\documentclass[12pt]{iopart}

\usepackage{iopams}
\usepackage{graphicx}

\begin{document} 

\title[]{Magnetization, magnetostriction, and \\ their relationship in Invar Fe$_{1-x}A_{x}$ ($A={\rm Pt},{\rm Ni}$)}

\author{Fran\c{c}ois Liot$^{1,2,3}$\footnote{Present address: Norinvar, 59 la rue, 50110 Bretteville, France.}}
\address{$^{1}$ Norinvar, 59 la rue, 50110 Bretteville, France}
\address{$^{2}$ Department of Computational Materials Design (CM), Max-Planck-Institut f{\"u}r Eisenforschung GmbH, D-40237 D{\"u}sseldorf, Germany}
\address{$^{3}$ Department of Physics, Chemistry, and Biology (IFM), Link{\"o}ping University, SE-581 83 Link{\"o}ping, Sweden}
\ead{f.liot@norinvar.com}

\begin{abstract}

A method is proposed for investigating the spontaneous magnetization, the spontaneous volume magnetostriction, and their relationship in disordered face-centered-cubic Fe$_{0.72}$Pt$_{0.28}$ and Fe$_{0.65}$Ni$_{0.35}$ in the temperature interval $0 \leq T/T_{\rm C} < 1$. It relies on the disordered local moment formalism and the observation that the reduced magnetization in each of the investigated materials is accurately described by an equation of the form $M(T)/M(0) = [ 1 -s (T/T_{\rm C} )^{3/2}- (1-s)(T/T_{\rm C})^{p} ]^{q}$. The present approach yields interesting results. The alloys at zero Kelvin share several physical properties: the volume in a partially disordered local moment state shrinks as the fraction of Fe moments which point down increases in the interval $0 < x^{{\rm Fe}\downarrow} < 1/2$, following closely $V(0) - 4 [V(0)-V(1/2)] x^{{\rm Fe}\downarrow} (1-x^{{\rm Fe}\downarrow})$, while the magnetization collapses, following closely $M(0) - 2 M(0) x^{{\rm Fe}\downarrow}$; the volume in the homogeneous ferromagnetic state greatly exceeds that in the disordered local moment state; $x^{{\rm Fe}\downarrow}(0)$ is close to zero. These common properties can account for a variety of intriguing phenomena displayed by both alloys, including the anomaly in the magnetostriction at zero Kelvin and, more surprisingly perhaps, the scaling between the reduced magnetostriction and the reduced magnetization squared below the Curie temperature. However, the thermal evolution of the fraction of Fe moments which point down depends strongly on the alloy under consideration. This, in turn, can explain the observed marked difference in the temperature dependence of the reduced magnetization between the two alloys.

\end{abstract}

\maketitle

\section{Introduction} \label{sec_intro}

Disordered face-centered-cubic (fcc) Fe$_{0.65}$Ni$_{0.35}$ and Fe$_{0.72}$Pt$_{0.28}$ alloys have received considerable attention due to their intriguing physical properties. For instance, their spontaneous volume magnetostriction, $w_{\rm s}=(V-V^{\rm PM})/V^{\rm PM}$, which measures the relative deviation of the equilibrium volume with respect to the volume in a paramagnetic state, is anomalously large at $T=0\,{\rm K}$ compared to a typical ferromagnet \cite{oomi81}. Furthermore, their reduced magnetostriction, $w_{\rm s}/w_{\rm s}(0)$, scales with the square of their reduced magnetization, $\left[M/M(0)\right]^{2}$, up to a temperature near the Curie temperature, $T_{\rm C}$ \cite{oomi81,crangle63,sumiyama76,sumiyama79}. This scaling presents a puzzle, as in Fe$_{0.65}$Ni$_{0.35}$, unlike in the other Invar alloy, the reduced magnetization exhibits an anomalous temperature dependence \cite{crangle63,sumiyama76}. The most fascinating example of these phenomena has long been the Invar effect: The linear thermal expansion coefficient of the ferromagnets, $\alpha=\left(1/a\right) \left(\partial a/\partial T\right)_{P}$, where $a$ denotes the lattice parameter, is anomalously small $\left[ \alpha(T) \ll 10^{-5}\,{\rm K}^{-1} \right]$ over a wide range of temperature \cite{guillaume97,kussmann50}. 

One of the greatest challenges in condensed matter theory today lies in understanding all of the abovementioned phenomena within one framework. 

Over the years, a consensus has emerged that the Invar effect is related to the magnetic properties of the systems in question. On the issue of which models are appropriate, however, opinions differ. One strand in the literature favours the so-called 2$\gamma$-state model, where the iron atoms can switch between two magnetic states with different atomic volumes as the temperature is raised \cite{weiss63}. This approach, however, seems incompatible with the results of M\"ossbauer \cite{ullrich84} and neutron experiments \cite{brown01}. Another approach based on {\it ab initio} density functional theory (DFT) emphasizes the importance of non-collinearity of the local magnetic moments on iron sites \cite{vanschilfgaarde99,dubrovinsky01}, though experiments undertaken to detect such non-collinearity have not found it \cite{cowlam03}. A third class of models relies on the disordered local moment (DLM) approach \cite{gyorffy85,johnson90,crisan02,khmelevskyi03,ruban07,liot12}, in which a binary alloy Fe$_{1-x}A_{x}$ with complete positional disorder of `up- and down-moments' on Fe sites is simulated, within the coherent potential approximation (CPA), as a three-component alloy Fe$^{\uparrow}_{(1-x)(1-x^{{\rm Fe}\downarrow})}$Fe$^{\downarrow}_{(1-x)x^{{\rm Fe}\downarrow}}A_{x}$. Here, $x^{{\rm Fe}\downarrow}$ represents the fraction of Fe moments which point down. 

\begin{figure}
\includegraphics[width=8cm]{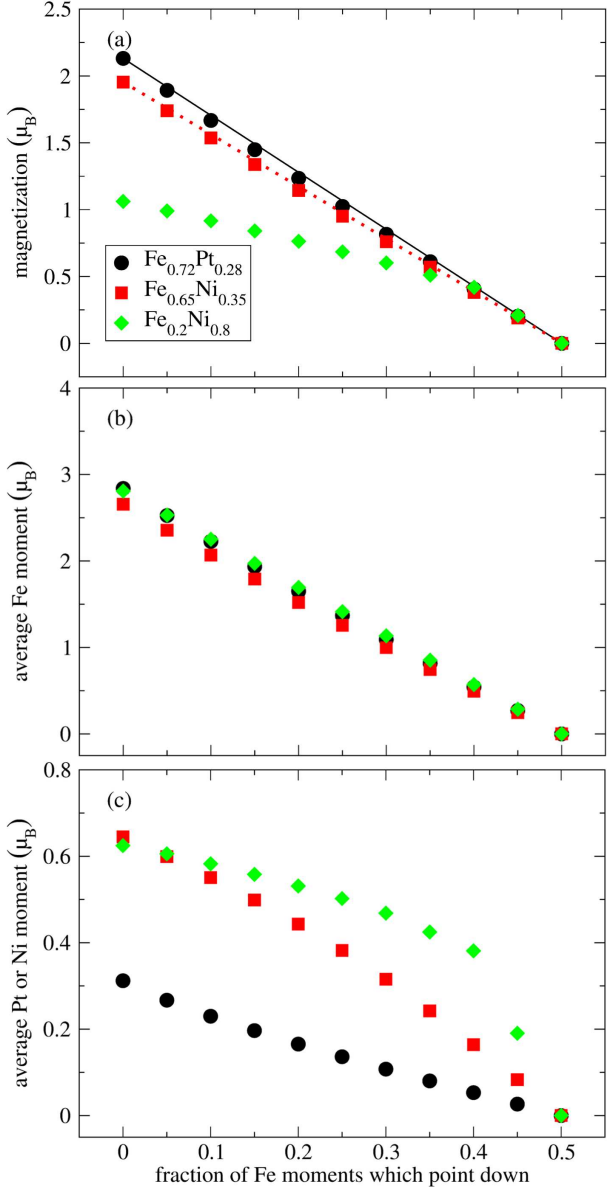}
\caption{The magnetization $M(x^{{\rm Fe}\downarrow})$ [panel~(a)], the average Fe moment $M^{\rm Fe}(x^{{\rm Fe}\downarrow})$ [panel~(b)], and the average {\it A} moment $M^{A}(x^{{\rm Fe}\downarrow})$ [panel~(c)] plotted against the fraction of Fe moments which point down. {\it A} refers to Pt or Ni, depending on the alloy under consideration. Symbols show results of {\it ab initio} calculations, which are performed at $T=0\,{\rm K}$. Lines correspond to polynoms of the form $M(0) - 2 M(0) x^{{\rm Fe}\downarrow}$.}
\label{figure2}
\end{figure}

In two recent exciting papers \cite{khmelevskyi03,liot12}, alloys in equilibrium at temperature $T$ in the range $0 \leq T/T_{\rm C} < 1$ have been modelled by random substitutional alloys in homogeneous ferromagnetic (FM) states, partially disordered local moment (PDLM) states, or DLM states depending on the fraction of Fe moments which are antiferromagnetically aligned with the spontaneous magnetization at $T$, $x^{{\rm Fe}\downarrow}(T)$. The general procedure can be divided into three stages. In the first stage, physical properties of interest (e.g., volume) are calculated for FM ($x^{{\rm Fe}\downarrow}=0$), PDLM ($0 < x^{{\rm Fe}\downarrow} < 1/2$), and DLM ($x^{{\rm Fe}\downarrow}=1/2$) states using {\it ab initio} DFT. In the second stage, the effect of thermal fluctuations on the fraction of Fe moments which point down, $x^{{\rm Fe}\downarrow}$, is investigated by means of a local moment model. Finally, combining the output from the two previous stages yields temperature-dependent properties. In \cite{khmelevskyi03}, the magnetization and the magnetic contribution to the fractional length change in Fe$_{0.7}$Pt$_{0.3}$ have been predicted. Even though simulation agrees qualitatively with experiment, the discrepancy between the calculated reduced magnetization at $T/T_{\rm C}=1/2$ and the corresponding experimental value exceeds 0.1. In \cite{liot12}, the linear thermal expansion coefficient in Fe$_{0.72}$Pt$_{0.28}$ and Fe$_{1-x}$Ni$_{x}$ with $x=0.35,0.4,\cdots,0.8$ has been investigated. The finding that Fe$_{0.72}$Pt$_{0.28}$ and Fe$_{0.65}$Ni$_{0.35}$ display the Invar effect perfectly matches experimental observation. Even the significant reduction of the thermal expansion coefficient in Fe$_{1-x}$Ni$_{x}$ at room temperature when $x$ is decreased from 0.55 to 0.35, which has been discovered by Guillaume \cite{guillaume21}, is reliably reproduced. It should be noted that, as in \cite{khmelevskyi03}, the approach developed in \cite{liot12} can fail to establish strong quantitative agreement with experiment. For instance, the measured structural quantity in Fe$_{0.65}$Ni$_{0.35}$ at $T=100\,{\rm K}$ is underestimated by more than $5\,10^{-6}\,{\rm K}^{-1}$. 

A possible source of discrepancies between simulation and experiment in \cite{khmelevskyi03,liot12}, along with PDLM and DLM states, may be associated with inaccurate results for the fraction of Fe moments which point down. In \cite{khmelevskyi03}, $x^{{\rm Fe}\downarrow}(T)$ in the temperature interval $0 \leq T/T_{\rm C} < 1$ has been calculated using a modified Weiss model, while an Ising (`up or down') model with nearest-neighbour interactions has been employed in \cite{liot12}. These two models probably offer the simplest ways of arriving at a value for $x^{{\rm Fe}\downarrow}(T)$. However, mean field theory may give a grossly inadequate result. This is especially true for random substitutional alloys where different atoms experience different chemical environments \cite{abrikosov07}. Furthermore, the Ising model ignores the possibility that Fe moments shrink with increasing temperature (see \cite{khmelevskyi03} and section~\ref{results}) and magnetic interactions remain non-negligible over a very long range \cite{pajda01,ruban05} due to their Ruderman-Kittel-Kasuya-Yoshida (RKKY) character.

\begin{figure}
\includegraphics[width=8cm]{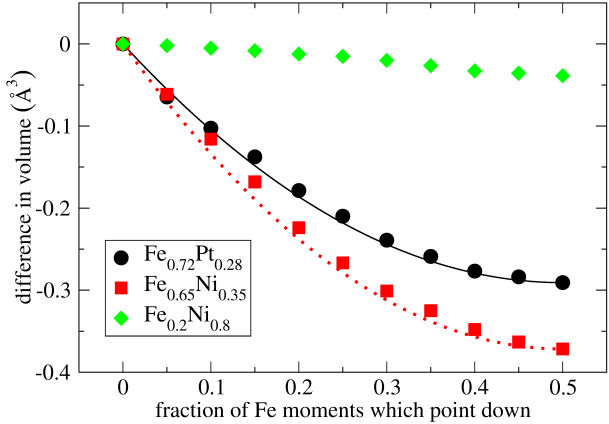}  
\caption{The difference in volume $[V(x^{{\rm Fe}\downarrow})-V(0)]$ plotted against the fraction of Fe moments which point down. Symbols show results of {\it ab initio} calculations, which are performed at $T=0\,{\rm K}$. Lines correspond to polynoms of the form $- 4 [V(0)-V(1/2)] x^{{\rm Fe}\downarrow} (1-x^{{\rm Fe}\downarrow})$. The reference quantities $V(0)$ are specified in table~\ref{table2}: $V(0)=13.44\,{\rm \AA}^{3}$ for Fe$_{0.72}$Pt$_{0.28}$ and $V(0)=11.59\,{\rm \AA}^{3}$ for Fe$_{0.65}$Ni$_{0.35}$.}
\label{figure1}
\end{figure}

This paper deals with the magnetization, the magnetostriction, and their relationship in Fe$_{0.72}$Pt$_{0.28}$ and Fe$_{0.65}$Ni$_{0.35}$ in the temperature interval $0 \leq T/T_{\rm C} < 1$. Taking a similar approach as in \cite{khmelevskyi03,liot12}, we model each of the investigated alloys in equilibrium at temperature $T$ by a random substitutional alloy in a FM, PDLM, or DLM state depending on $x^{{\rm Fe}\downarrow}(T)$. While the local moment models employed in \cite{khmelevskyi03,liot12} suffer from drawbacks, predicting correctly the magnetic ground state of Fe-Pt and Fe-Ni alloys using first-principles calculations based on local exchange-correlation functionals has so far proven impossible \cite{abrikosov07}, and classical spin dynamics turns out inadequate to describe finite-temperature magnetic properties of ferromagnets \cite{kuzmin05}, we predict $x^{{\rm Fe}\downarrow}(T)$ from the observation that the reduced magnetization is accurately described by an equation of the form 
\small
\begin{eqnarray}\label{eqn4}
\frac{M(T)}{M(0)} = \bigg[ 1 -s \bigg( \frac{T}{T_{\rm C}} \bigg)^{3/2}- (1-s) \bigg( \frac{T}{T_{\rm C}} \bigg)^{p} \bigg]^{q}.
\end{eqnarray}
\normalsize

Section~\ref{comp_methods} is devoted to computational details. Section~\ref{results} presents a comprehensive discussion of our results.

\section{Computational methodology}\label{comp_methods}

Having thus outlined the general approach, we now present the details, with some commentary.

As a first step, we perform calculations of the magnetization and the volume at $T=0\,{\rm K}$ in FM, PDLM, and DLM states. This is done within the framework of the exact muffin-tin orbitals (EMTO) theory in combination with the full charge density (FCD) technique \cite{vitos01} and using the generalized gradient approximation (GGA) \cite{perdew96}. Static ionic displacements are neglected \cite{liot06,liot09,liot_thesis}. As in recent theoretical studies on Fe-Pt \cite{khmelevskyi03,liot12, khmelevskyi03-PRB} and Fe-Ni \cite{ruban07,liot12,abrikosov07,ekholm10}, complete positional disorder of chemical species on fcc lattice sites and up- and down-moments on Fe sites is treated within the CPA \cite{vitos-abrikosov01}. A Monkhorst-Pack grid of $41 \times 41 \times 41$ $\mathbf{k}$-points \cite{monkhorst76} is chosen to ensure convergence of the volume $V(x^{{\rm Fe}\downarrow})$ and the magnetization $M(x^{{\rm Fe}\downarrow})$ to better than $10^{-2}\,{\rm \AA}^{3}$ and $10^{-2}\,\mu_{\rm B}$, respectively.

\begin{figure}
\includegraphics[width=8cm]{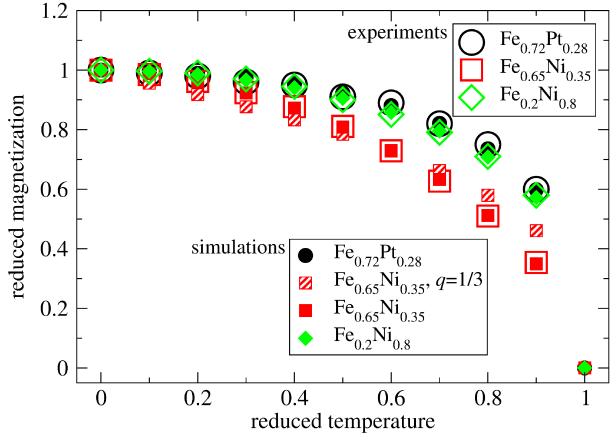}
\caption{The reduced magnetization $M(T)/M(0)$ plotted against the reduced temperature $T/T_{\rm C}$. Open symbols show experimental data \cite{crangle63,sumiyama76}. Filled and hatched symbols correspond to equation~(\ref{eqn4}).}
\label{figure3}
\end{figure}

\begin{figure}
\includegraphics[width=8cm]{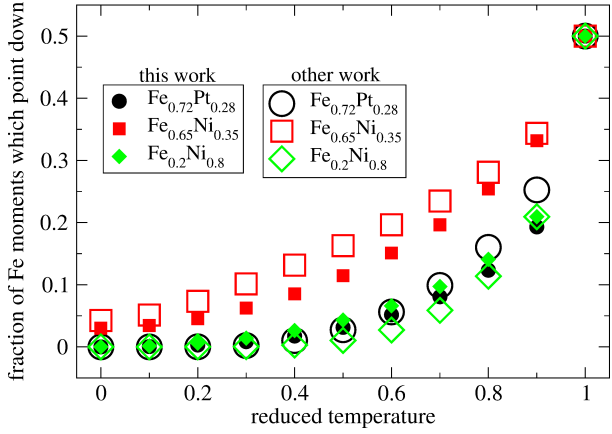}
\caption{The fraction of Fe moments which are antiferromagnetically aligned with the magnetization at temperature $T$, $x^{{\rm Fe}\downarrow}(T)$, plotted against the reduced temperature $T/T_{\rm C}$, according to this work and a previous study \cite{liot12}.}
\label{figure4}
\end{figure}

In the second step, we turn to address the thermal evolution of the fraction of Fe moments which point down and proceed as follows. First, we observe that an accurate description of the reduced magnetization is provided by (\ref{eqn4}), where $p$ and $q$ are parameters, $p>3/2$. In the definition
\small
\begin{eqnarray}\label{eqn3}
s = \frac{3}{8} \pi^{-3/2} \zeta(3/2) V(0) \frac{\mu_{B}}{M(0)} \bigg[ \frac{k_{\rm B} T_{\rm C}}{D(0)} \bigg]^{3/2},
\end{eqnarray}
\normalsize
which relies on classical spin-wave theory \cite{kittel96}, $\zeta$ stands for Riemann's zeta function and $D(0)$ denotes the zero-temperature spin-wave stiffness. Second, we assume
\small
\begin{eqnarray}\label{eqn5}
x^{{\rm Fe}\downarrow}(T) = \frac{1}{2} - \bigg[ \frac{1}{2} - x^{{\rm Fe}\downarrow}(0) \bigg] \bigg[ 1- \left( \frac{T}{T_{\rm C}} \right)^{p} \bigg]^{q}.
\end{eqnarray}
\normalsize
Note that realizing this scheme requires the prior knowledge of the reduced magnetization as a function of the reduced temperature, the dimensionless quantity $s$, and the fraction of Fe moments which point down at $T=0\,{\rm K}$.
  
Results, which can be found in this paper and recent theoretical studies \cite{khmelevskyi03,liot12}, provide some physical justification for the abovementioned hypothesis. As shown in section~\ref{results}, the zero-temperature magnetization $M(x^{{\rm Fe}\downarrow})$ decays linearly with increasing $x^{{\rm Fe}\downarrow}$ in the range $0 \leq x^{{\rm Fe}\downarrow} \leq 1/2$. Thus, the specific power law behaviour displayed by the reduced magnetization [see (\ref{eqn4})] may reflect the thermal evolution of $x^{{\rm Fe}\downarrow}$. However, (\ref{eqn4}) is constructed to obey Bloch's 3/2 power law at low temperatures and we expect the term proportional to $(T/T_{\rm C} )^{3/2}$ in this equation to be related to spin-wave excitations. We therefore deliberately exclude such a term from (\ref{eqn5}). Moreover, according to the Ising model \cite{liot12}, $x^{{\rm Fe}\downarrow}$ increases continuously upon heating the system and, in the limit $T/T_{\rm C} \rightarrow 1^{-}$, $x^{{\rm Fe}\downarrow}(T) \rightarrow 1/2$. This is exactly what we expect to find for Fe$_{0.72}$Pt$_{0.28}$ from the modified Weiss model \cite{khmelevskyi03}. Combining the above findings leads to the simple relationship~(\ref{eqn5}). 

In the third and final step, we combine the output from the previous steps to explore how the magnetization $M [x^{{\rm Fe}\downarrow}(T)]$ and the magnetostriction 
\small
\begin{eqnarray}\label{eqn100}
w_{\rm s}[x^{{\rm Fe}\downarrow}(T)] = \frac{V [x^{{\rm Fe}\downarrow}(T)] - V (1/2)}{V(1/2)}.
\end{eqnarray}
\normalsize
vary as the system is heated. Note that $w_{\rm s}[x^{{\rm Fe}\downarrow}(T)]$ measures the relative change of the volume upon lowering the fraction of Fe moments which point down from 1/2 to $x^{{\rm Fe}\downarrow}(T)$. At $T=0\,{\rm K}$, in the case of a homogeneous ferromagnet, it coincides with the expression $[V (0) - V (1/2)]/V(1/2)$ proposed in \cite{khmelevskyi03,ruban07,khmelevskyi04}.

\section{Results and discussion}\label{results}

\subsection{Physical properties at $T=0\,{\rm K}$ for FM, PDLM, and DLM states}\label{results A}

Figure~\ref{figure2} displays the calculated magnetization $M(x^{{\rm Fe}\downarrow})$ [panel~(a)], the calculated average Fe moment $M^{\rm Fe}(x^{{\rm Fe}\downarrow})$ [panel~(b)], and the calculated average {\it A} moment $M^{A}(x^{{\rm Fe}\downarrow})$ [panel~(c)] in Fe$_{0.72}$Pt$_{0.28}$ and Fe$_{0.65}$Ni$_{0.35}$ at $T=0\,{\rm K}$ for FM, PDLM, and DLM states. {\it A} refers to Pt or Ni, depending on the alloy under consideration. The value of $M(0)$ is reported in table~\ref{table2} \cite{oomi81,crangle63,sumiyama76,sumiyama79,acet94,hayase73}. Moving away from $x^{{\rm Fe}\downarrow}=0$, the magnetization decreases rapidly, following closely
\small
\begin{eqnarray}\label{eqn2}
M(0) - 2 M(0) x^{{\rm Fe}\downarrow};
\end{eqnarray}
\normalsize
it eventually cancels for $x^{{\rm Fe}\downarrow}=1/2$. This result has lead us to assume that the fraction of Fe moments which point down is given by (\ref{eqn5}). The magnetization relates to the average Fe moment through
\small
\begin{eqnarray}\label{eqn6}
M(x^{{\rm Fe}\downarrow}) = (1-x) M^{\rm Fe}(x^{{\rm Fe}\downarrow}) + xM^{A}(x^{{\rm Fe}\downarrow}).
\end{eqnarray}
\normalsize
Neither Fe$_{0.72}$Pt$_{0.28}$ nor Fe$_{0.65}$Ni$_{0.35}$ exhibits noticeable anomalies in their average Fe moment. This may come as a surprise since raising $x^{{\rm Fe}\downarrow}$ from 0 to 1/2 causes, for example, a drastic reduction (up to 16\%) of the average Fe moment which point up.

Figure~\ref{figure1} shows the calculated volume difference $\Delta V(x^{{\rm Fe}\downarrow})=[V(x^{{\rm Fe}\downarrow})-V(0)]$ in the alloys at $T=0\,{\rm K}$ for FM, PDLM, and DLM states. Table~\ref{table2} displays the value of $V(0)$. The volume $V$ shrinks continously with increasing $x^{{\rm Fe}\downarrow}$, behaving in a similar way to 
\small
\begin{eqnarray}\label{eqn1}
V(0)- 4 [V(0)-V(1/2)] x^{{\rm Fe}\downarrow} (1-x^{{\rm Fe}\downarrow}).
\end{eqnarray}
\normalsize
Section~\ref{results C} provides evidence that this feature is linked with the scaling between the reduced magnetostriction and the reduced magnetization squared, which has been observed experimentally below the Curie temperature. Such a property is bound to exist in other materials also, as it has been detected in Fe$_{0.2}$Ni$_{0.8}$ [$\Delta V(1/2)=-0.04\,{\rm \AA}^{3}$], which does not exhibit any major anomalies in thermal expansion, and Fe$_{0.8}$Ni$_{0.2}$ [$\Delta V(1/2)=-0.68\,{\rm \AA}^{3}$], which shows the anti-Invar effect. We now turn to the volume difference between the DLM state and the FM state. The calculated $\Delta V(1/2)$ amounts to $-0.29\,{\rm \AA}^{3}$ in Fe$_{0.72}$Pt$_{0.28}$ and $-0.37\,{\rm \AA}^{3}$ in Fe$_{0.65}$Ni$_{0.35}$. These numerical data agree with previous estimates \cite{khmelevskyi03,ruban07}. Contrary to Fe$_{0.2}$Ni$_{0.8}$, $\Delta V(1/2)$ in these two ferromagnets is sufficiently large to compensate for the volume expansion in the corresponding paramagnets due to heating from $0\,{\rm K}$ to $T_{\rm C}$ \cite{oomi81,tanji71}. In this context, an understanding of the circumstances that give rise to a strong dependence of $V$ on $x^{{\rm Fe}\downarrow}$ can play an important role in the improvement of our knowledge of the Invar effect \cite{liot-hooley14}. At this stage, we infer that the volume dependence of exchange parameters of a classical Heisenberg Hamiltonian influences how the volume $V$ varies with $x^{{\rm Fe}\downarrow}$ \cite{liot14}.

\subsection{The fraction of Fe moments which point down at $T \geq 0\,{\rm K}$}\label{results B}

\begin{table}
\caption{\label{table2} Theoretical predictions for magnetizations, volumes, and magnetostrictions at $T=0\,{\rm K}$ compared with experimental measurements below $T=5\,{\rm K}$ \cite{oomi81,crangle63,sumiyama76,sumiyama79,acet94,hayase73}. Note that $x^{{\rm Fe}\downarrow}=0$ corresponds to homogeneous ferromagnetic states and $0 < x^{{\rm Fe}\downarrow} < 1/2$ to PDLM states.}
\begin{indented}
\item[]\begin{tabular}{@{}*{5}{c}}
\br
& & magnetization & volume & magnetostriction \\
alloy & reference & ($\mu_{\rm B}$) &  \big(\AA$^{3}$\big) &  \big($10^{-2}$\big) \\ 
\mr
Fe$_{0.72}$Pt$_{0.28}$ & simulations, $x^{{\rm Fe}\downarrow}=0$ & 2.13 & 13.44 & 2.21 \\
                           & experiments  & 2.09 & 13.14 & 1.52-1.60 \\
Fe$_{0.65}$Ni$_{0.35}$ & simulations, $x^{{\rm Fe}\downarrow}=0$ & 1.95 & 11.59 & 3.31 \\
                           & simulations, $x^{{\rm Fe}\downarrow}=0.03$ & 1.82 & 11.55 & 2.99 \\
                           & experiments & 1.78 & 11.61 & 1.79-2.2 \\
Fe$_{0.2}$Ni$_{0.8}$ & simulations, $x^{{\rm Fe}\downarrow}=0$ & 1.06 & 11.13 & 0.35 \\
                         & experiments & 1.06 & 11.08 & 0 \\
\br
\end{tabular}
\end{indented}
\end{table}

\begin{table}
\caption{\label{table1} Experimental data for the Curie temperature and a spin-wave stiffness obtained below $T=5\,{\rm K}$ \cite{crangle63,sumiyama76,sumiyama79,acet94,grigoriev02,hennion75,rosov94} together with the calculated physical constant $s$ [equation~(\ref{eqn3})] and the fitting parameters $p$ and $q$ [equation~(\ref{eqn4})].}
\begin{indented}
\item[]\begin{tabular}{@{}*{6}{c}}
\br
& Curie temperature & spin-wave stiffness &  \\ 
alloy &  (K) &  (meV\,\AA$^{2}$) & $s$ & $p$ & $q$ \\ 
\mr
Fe$_{0.72}$Pt$_{0.28}$ & 370 & 98 & 0.205 & 5/2 & 1/3\\
Fe$_{0.65}$Ni$_{0.35}$ & 500 & 117 & 0.26 & 1.8 & 0.585\\
Fe$_{0.2}$Ni$_{0.8}$ & 840 & 336 & 0.18 & 2.075 & 1/3\\
\br
\end{tabular}
\end{indented}
\end{table}

The fraction of Fe moments which point down is an important quantity for physical insight and further theoretical analysis. To investigate its thermal evolution in Fe$_{0.72}$Pt$_{0.28}$ and Fe$_{0.65}$Ni$_{0.35}$, we use a scheme described in section~\ref{comp_methods}. 

To begin with, we collect reliable experimental data. Figure~\ref{figure3} reports on extensive magnetization measurements \cite{crangle63,sumiyama76}. Table~\ref{table1} provides the physical constants $s$ \cite{crangle63,sumiyama76,sumiyama79,acet94,grigoriev02,hennion75,rosov94}, which have been estimated by exploiting inelastic neutron scattering \cite{grigoriev02,hennion75,rosov94}. Furthermore, a literature survey yields $x^{{\rm Fe}\downarrow}(0)=0$ for Fe$_{0.72}$Pt$_{0.28}$ \cite{nakamura79} and $x^{{\rm Fe}\downarrow}(0) \approx 3$\% for Fe$_{0.65}$Ni$_{0.35}$ \cite{abd87}.

Second, we fit experimental data for the reduced magnetization with the analytic expression (\ref{eqn4}). As shown in figure~\ref{figure3}, it turns out that both alloys comply with (\ref{eqn4}). We are unclear as to why this is the case. Nonetheless, this finding represents a breakthrough in magnetism \cite{wassermann90}. The fit parameters $p$ and $q$ are displayed in table~\ref{table1}. For Fe$_{0.72}$Pt$_{0.28}$, $s=0.205$, $p=5/2$, and $q=1/3$. The temperature dependence of the reduced magnetization of this alloy resembles that of other metallic ferromagnets such as fcc Ni and fcc Co \cite{kuzmin05} to name just a few. Switching to Fe$_{0.65}$Ni$_{0.35}$ leads to a 76\% enhancement of the parameter $q$ and thus anomalous reduced magnetization.

We then combine the values for $x^{{\rm Fe}\downarrow}(0)$, $p$, and $q$ given above with (\ref{eqn5}). The corresponding results are reported in figure~\ref{figure4}, together with earlier data \cite{liot12}. Raising temperature affects the magnetic phase of both alloys, shifting $x^{{\rm Fe}\downarrow}(T)$ to higher values. A crucial feature of this description is the strong excess of $x^{{\rm Fe}\downarrow}(T)$ in Fe$_{0.65}$Ni$_{0.35}$ with respect to Fe$_{0.72}$Pt$_{0.28}$ over a wide range of temperature. For instance at $T/T_{\rm C}=1/2$, $x^{{\rm Fe}\downarrow}(T)$ amounts to 11\% in the former but only 3\% in the latter. We note the striking resemblance between this physical situation and the picture provided by the highly oversimplified Ising model.

\subsection{Other physical properties at $T \geq 0\,{\rm K}$}\label{results C}

\begin{figure}
\includegraphics[width=8cm]{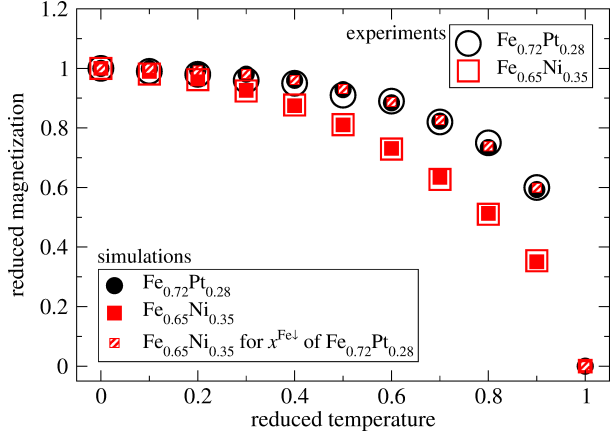}
\caption{The reduced magnetization according to simulations and experiments \cite{crangle63,sumiyama76} plotted against the reduced temperature.}
\label{figure5}
\end{figure}

\begin{figure}
\includegraphics[width=8cm]{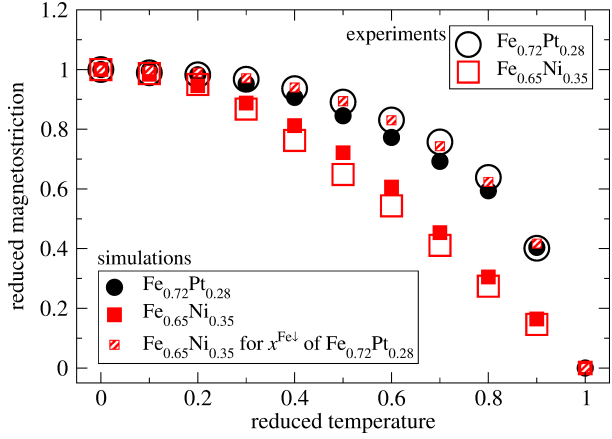}
\caption{The analog of figure~\ref{figure5} for the reduced magnetostriction. Experimental data are taken from \cite{oomi81,sumiyama79}.}
\label{figure6}
\end{figure}

Our results for the magnetization and the magnetostriction at $T=0\,{\rm K}$ are reported in Table~\ref{table2}, along with experimental data. Our simulations give $M [x^{{\rm Fe}\downarrow}(0)]=2.13\,\mu_{\rm B}$ in Fe$_{0.72}$Pt$_{0.28}$ and $M [x^{{\rm Fe}\downarrow}(0)]=1.82\,\mu_{\rm B}$ in Fe$_{0.65}$Ni$_{0.35}$. These values differ only marginally by less than 3\% from experimental evidence \cite{crangle63,sumiyama76}. Considering the high sensitivity of $M$ to $x^{{\rm Fe}\downarrow}$ [figure~\ref{figure2}(a)], this agreement is remarkable. Furthermore, we find that $w_{\rm s} [x^{{\rm Fe}\downarrow}(0)]$ amounts to $2.21\,10^{-2}$ in the former and $2.99\,10^{-2}$ in the latter. As expected from earlier work \cite{khmelevskyi03,ruban07}, the theoretical values systematically overestimate the experimental ones \cite{oomi81,sumiyama79,hayase73}. In contrast to \cite{ruban07}, our approach applied to Fe$_{0.65}$Ni$_{0.35}$ incorporates the deviation of $x^{{\rm Fe}\downarrow}(0)$ from 0. This implementation reduces the discrepancy with measurements by at least 25\% \cite{oomi81,hayase73}.   

Figure~\ref{figure5} displays the reduced magnetization in the alloys at various temperatures in the range $0 \leq T/T_{\rm C} < 1$, as obtained from simulations and experiments. Moving away from $T/T_{\rm C}=0$, the calculated quantity of interest falls regardless of the nature of the system, with the largest decay occuring in Fe$_{0.65}$Ni$_{0.35}$. Theoretical predictions reproduce accurately measurements \cite{crangle63,sumiyama76}, in particular the anomalously low values for Fe$_{0.65}$Ni$_{0.35}$ at $T/T_{\rm C}=0.5,0.6,\cdots,0.9$.

Figure~\ref{figure6} is the analog of figure~\ref{figure5} for the reduced magnetostriction. Features in the structural data resemble those seen in the reduced magnetization results. The agreement between simulations and experiments \cite{oomi81,sumiyama79} is satisfactory, although it deteriorates when our attention shifts from figure~\ref{figure5} to~\ref{figure6}.

\begin{figure}
\includegraphics[width=8cm]{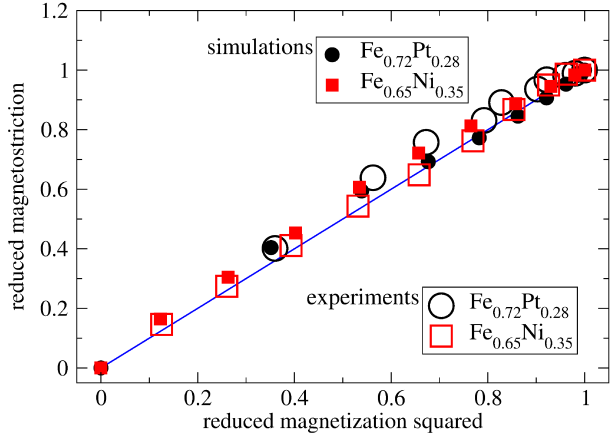}
\caption{Magnetostructural coupling below the Curie temperature.}
\label{figure7}
\end{figure}

We study in figure~\ref{figure7} the relationship between the reduced magnetostriction and the reduced magnetization squared below $T_{\rm C}$. Our resulting data points collapse fairly well onto the straight line passing through the origin and with a slope of one. They lie close to their corresponding experimental estimates, as expected from the two previous figures. 

Judging from table~\ref{table2} and figures~\ref{figure5},~\ref{figure6}, and~\ref{figure7}, the formalism developed in section~\ref{comp_methods} achieves a good description of magnetic, structural, and magnetostructural properties of Fe$_{0.72}$Pt$_{0.28}$ and Fe$_{0.65}$Ni$_{0.35}$. Unlike the computational methodology in \cite{khmelevskyi03}, it can be applied with success to Fe-Ni alloys. In addition, evidence suggests that it yields more precise results for Fe-Pt. The validation of our approach opens exciting opportunities for investigating the mechanism of intriguing phenomena, which, in principle, can now be understood within the same framework. 

\subsection{Origin of observed phenomena}\label{results D}

To begin with, we consider the magnetostriction in Fe$_{0.72}$Pt$_{0.28}$ and Fe$_{0.65}$Ni$_{0.35}$ at $T=0\,{\rm K}$. A natural question to ask is: What is the origin of the anomalously large values observed experimentally? To explore this within our theoretical framework, we point out that
\small
\begin{eqnarray}\label{eqn7}
\fl w_{\rm s} [x^{{\rm Fe}\downarrow}(0)] = \frac{V(0)-V(1/2)}{V(1/2)} 4 [1/2 - x^{{\rm Fe}\downarrow}(0)]^{2} \{ 1 + \epsilon_{\rm vol}[x^{{\rm Fe}\downarrow}(0)] \}, 
\end{eqnarray}
\normalsize
where
\normalsize
\small
\begin{eqnarray}\label{eqn7_1}
\fl \epsilon_{\rm vol} (x^{{\rm Fe}\downarrow}) = \nu_{\rm vol}(x^{{\rm Fe}\downarrow}) / 4 (1/2 - x^{{\rm Fe}\downarrow})^{2} \nonumber
\end{eqnarray}
\normalsize
and
\normalsize
\small
\begin{eqnarray}\label{eqn7_2}
\fl \nu_{\rm vol} (x^{{\rm Fe}\downarrow}) = \{ V ( x^{{\rm Fe}\downarrow} ) - V ( 0 )
+ 4 [ V ( 0 ) - V (1/2) ] x^{{\rm Fe}\downarrow} ( 1-x^{{\rm Fe}\downarrow} ) \} / [ V ( 0 ) - V (1/2) ]. \nonumber
\end{eqnarray}
\normalsize
Consequently, $w_{\rm s} [x^{{\rm Fe}\downarrow}(0)]$ can be written as the product of three terms: $[V(0)-V(1/2)]/V(1/2)$, $4 [1/2 - x^{{\rm Fe}\downarrow}(0)]^{2}$, and $\{ 1 + \epsilon_{\rm vol}[x^{{\rm Fe}\downarrow}(0)] \}$. The first term denotes the magnetostriction that would exhibit the investigated material if it were FM. It has been studied previously \cite{khmelevskyi03,ruban07,khmelevskyi04} and its strong value in Fe$_{0.72}$Pt$_{0.28}$ has been linked to the substantial change undergone by the magnitude of Fe moments on switching the magnetic state from DLM to FM \cite{khmelevskyi04}. The product of the two first terms represents the magnetostriction that would display the material if its volume $V(x^{{\rm Fe}\downarrow})$ were given by (\ref{eqn1}) in $0 < x^{{\rm Fe}\downarrow} < 1/2$. The second term when multiplied by the third one decreases steeply with increasing $x^{{\rm Fe}\downarrow}(0)$ in the range 0-0.4. We note in passing that $x^{{\rm Fe}\downarrow}(0)$ may reach high values in Fe-rich Fe$_{1-x}$Ni$_{x}$ such as 0.33 for $x=0.3$ \cite{rancourt96}. Based on this analysis, we claim that the anomalies can be traced back to the combination of two zero-temperature properties: the volume in the FM state greatly exceeds that in the DLM state and the fraction of Fe moments which point down is close to 0.  

Over the last few decades, several arguments have been put forward to justify the unusual thermal evolution of the reduced magnetization reported in some Invar systems, including anomalous spin-wave damping mechanism \cite{ishikawa81} and anomalous average magnitude of Fe moments \cite{khmelevskyi03}. However, high-resolution inelastic neutron scattering data contradict the first argument (see, e.g., \cite{kaul94} and references therein), while the second scenario is difficult to reconcile with the absence of noticeable anomaly in the $x^{{\rm Fe}\downarrow}$ dependence of the average Fe moment, $M^{\rm Fe}$ [figure~\ref{figure2}(b)]. In this regard, we emphasize that $M^{\rm Fe}$ represents the only contribution to $M$ associated with Fe sites [equation~(\ref{eqn6})]. To advance the matter further, we plot in figure~\ref{figure5} the calculated reduced magnetization that would exhibit Fe$_{0.65}$Ni$_{0.35}$ if its fraction of Fe moments which point down matched that of Fe$_{0.72}$Pt$_{0.28}$. In this case, the data for the former alloy mimic the behaviour of the latter. This substantiates our claim that the anomaly detected in the reduced magnetization in Fe$_{0.65}$Ni$_{0.35}$ arises from the peculiar thermal evolution of the fraction of Fe moments which point down.

Figure~\ref{figure6} is the analog of figure~\ref{figure5} for the reduced magnetostriction. It reveals that if the fraction of Fe moments which point down of Fe$_{0.65}$Ni$_{0.35}$ were substituted by that of Fe$_{0.72}$Pt$_{0.28}$ (see hatched symbols), the data for the former alloy would follow closely the behaviour of the latter. Thus, the thermal evolution of the fraction of Fe moments which point down can account for the observed marked difference in the temperature dependence of the reduced magnetostriction between the two alloys.

The reduced magnetization in Fe$_{0.65}$Ni$_{0.35}$ exhibits an anomalous temperature dependence, unlike in Fe$_{0.72}$Pt$_{0.28}$. For this reason, experimental evidence that the reduced magnetostriction decreases proportionally to the square of the reduced magnetization in each of these alloys as the system under study is heated from $T=0\,{\rm K}$ up to a temperature near $T_{\rm C}$ may come as a surprise. Explaining these results could help unravel the mechanism behind magnetostructural coupling in transition metals and their alloys. We propose below the first investigation of the origin of these remarkable phenomena using the DFT. We first note that
\small
\begin{eqnarray}\label{eqn8}
\fl \frac{ w_{\rm s} [x^{{\rm Fe}\downarrow}(T)] }{ w_{\rm s}[x^{{\rm Fe}\downarrow}(0)] } = \frac{ M^{2} [x^{{\rm Fe}\downarrow}(T)] }{ M^{2} [x^{{\rm Fe}\downarrow}(0)] } \lambda \mu \{ 1 + \epsilon[x^{{\rm Fe}\downarrow}(T)] \},
\end{eqnarray}
\normalsize
where
\small
\begin{eqnarray}\label{eqn9}
\fl \lambda = \frac{ w_{\rm s} (0) }{ w_{\rm s}[x^{{\rm Fe}\downarrow}(0)] }, \nonumber
\end{eqnarray}
\normalsize
\small
\begin{eqnarray}\label{eqn10}
\fl \mu = \frac{ M^{2} [x^{{\rm Fe}\downarrow}(0)] }{ M^{2}(0) }, \nonumber
\end{eqnarray}
\normalsize
\small
\begin{eqnarray}\label{eqn11}
\fl \epsilon(x^{{\rm Fe}\downarrow}) = \nu(x^{{\rm Fe}\downarrow}) \frac{ M^{2}(0) }{M^{2} (x^{{\rm Fe}\downarrow})}, \nonumber
\end{eqnarray}
\normalsize
\small
\begin{eqnarray}\label{eqn11}
\fl \nu(x^{{\rm Fe}\downarrow}) = \nu_{\rm vol} (x^{{\rm Fe}\downarrow})- 2 (1-2 x^{{\rm Fe}\downarrow}) \nu_{\rm mag} (x^{{\rm Fe}\downarrow}) - \nu_{\rm mag}^{2} (x^{{\rm Fe}\downarrow}), \nonumber
\end{eqnarray}
\normalsize
and
\small
\begin{eqnarray}\label{eqn13}
\fl \nu_{\rm mag} (x^{{\rm Fe}\downarrow}) = [ M(x^{{\rm Fe}\downarrow})-M(0)+2 M(0) x^{{\rm Fe}\downarrow}]/M(0). \nonumber
\end{eqnarray}
\normalsize
Thus if $x^{{\rm Fe}\downarrow}(0)$ were zero, $M(x^{{\rm Fe}\downarrow})$ were given by (\ref{eqn2}), and $V(x^{{\rm Fe}\downarrow})$ were given by (\ref{eqn1}) in $0 < x^{{\rm Fe}\downarrow} < 1/2$, then the reduced magnetostriction would match the reduced magnetization squared over the whole range $0 \leq T/T_{\rm C} < 1$. Taking into account this finding, our \emph{ab initio} results for $M$ and $V$ displayed in figures~\ref{figure2} and~\ref{figure1}, and the experimental data for $x^{{\rm Fe}\downarrow}(0)$ mentioned in section~\ref{results B}, we argue that common features observed experimentally in the relationship between the magnetization and the magnetostriction in Fe$_{0.72}$Pt$_{0.28}$ and Fe$_{0.65}$Ni$_{0.35}$ below their Curie temperature can originate from the fact that both alloys exhibit similar \emph{zero-temperature} properties: their magnetization and their volume in a PDLM state follow closely (\ref{eqn2}) and (\ref{eqn1}) and their fraction of Fe moments which point down is close to 0. 

\section{Conclusion} \label{conclusion}

To address the magnetization, the magnetostriction, and their relationship in disordered fcc Fe$_{0.72}$Pt$_{0.28}$ and Fe$_{0.65}$Ni$_{0.35}$ in the temperature interval $0 \leq T/T_{\rm C} < 1$, we develop a method in which each of the alloys in equilibrium at temperature $T$ is modelled by a random substitutional alloy in a FM, PDLM, or DLM state depending on $x^{{\rm Fe}\downarrow}(T)$. The method consists of three stages.

As a first step, we perform DFT calculations of the magnetization and the volume at $T=0\,{\rm K}$ in FM, PDLM, and DLM states. In the second step, we turn to the thermal evolution of the fraction of Fe moments which point down. To achieve this goal, we rely on the fact that an accurate description of the reduced magnetization is provided by (\ref{eqn4}). We also assume that the function $x^{{\rm Fe}\downarrow}$ obeys (\ref{eqn5}). In the third and final step, we combine the output from the previous steps to explore how the magnetization and the magnetostriction vary as the system is heated.

Our method appears to us sufficiently robust so that our following conclusions will remain unaffected. The alloys at $T=0\,{\rm K}$ share several physical properties: the magnetization in a PDLM state collapses as the fraction of Fe moments which point down increases, following closely (\ref{eqn2}), while the volume shrinks, following closely (\ref{eqn1}); the volume in the FM state greatly exceeds that in the DLM state; $x^{{\rm Fe}\downarrow}(0)$ is close to 0. These common properties can account for a variety of intriguing phenomena displayed by both alloys, including the anomaly in the magnetostriction at $T=0\,{\rm K}$ and, more surprisingly perhaps, the scaling between the reduced magnetostriction and the reduced magnetization squared below the Curie temperature. However, the thermal evolution of the fraction of Fe moments which point down depends strongly on the alloy under consideration. This, in turn, can explain the observed marked difference in the temperature dependence of the reduced magnetization between the two alloys.

\ack
The author thanks I. A. Abrikosov (Link{\"o}ping), B. Alling (Link{\"o}ping), C. A. Hooley (St Andrews, U.K.), A. E. Kissavos (Link{\"o}ping), and J. Neugebauer (D{\"u}sseldorf) for fruitful discussions.


\end{document}